\documentclass[twocolumn,aps,prl,showpacs,preprintnumbers,amsfonts,amsmath,amssymb,superscriptaddress,floatfix]
{revtex4}
\usepackage[dvipdfm]{graphicx}

\newcommand{\etal}{{\it et~al.\ }}
\newcommand{\beq}{\begin{equation}}
\newcommand{\eeq}{\end{equation}}
\newcommand{\beqa}{\begin{eqnarray}}
\newcommand{\eeqa}{\end{eqnarray}}
\newcommand{\beqan}{\begin{eqnarray*}}
\newcommand{\eeqan}{\end{eqnarray*}}

\newcommand{\abs}[1]{\left\vert #1 \right\vert}

\newcommand{\ket}[1]{\left\vert #1 \right\rangle}

\newcommand{\makebib}[5]{#1 #2 \textbf{#3}, #4 (#5).}
\newcommand{\refeq}[1]{Eq.~(\ref{#1})}
\newcommand{\reffig}[1]{Fig.~\ref{#1}}
\newcommand{\ani}{\hat{a}}

\let\eps=\epsilon 
\newcommand{\affA}{%
\affiliation{
     National Institute of Information and Communications Technology
     (NICT), \\
     4-2-1 Nukui-kitamachi, Koganei, Tokyo 184-8795, Japan}
     }
\newcommand{\affB}{%
\affiliation{
	Department of Applied Physics,
	School of Engineering,
	The University of Tokyo,\\
	7-3-1 Hongo Bunkyo-ku, Tokyo 113-8656, Japan}
	}

\newcommand{\affC}{%
\affiliation{
     CREST, Japan Science and Technology Agency, 
     5 Sanbancho, Chiyoda-ku, Tokyo 102-0075, Japan}
     }

\begin{document}
\title{Generation of large-amplitude coherent-state superposition 
via ancilla-assisted photon-subtraction}

\date{\today}
\author{Hiroki Takahashi}
\affA
\affB
\affC

\author{Kentaro Wakui}
\affA
\affC
\author{Shigenari Suzuki}
\author{Masahiro Takeoka}
\author{Kazuhiro Hayasaka}
\affA
\affC
\author{Akira Furusawa}
\affB
\affC
\author{Masahide Sasaki}%
\affA%
\affC%

\begin{abstract}
We propose and demonstrate a novel method to generate 
a large-amplitude coherent-state superposition (CSS)
via ancilla-assisted photon-subtraction. 
The ancillary mode induces quantum interference of 
indistinguishable processes in an extended space, widening the controllability 
of quantum superposition at the conditional output. 
We demonstrate this by a simple time-separated two-photon subtraction from continuous wave squeezed light. 
We observe the largest CSS of travelling light ever reported 
without correcting any imperfections, which will enable various quantum 
information applications with CSS states. 
\end{abstract}



\maketitle

Superposition of macroscopically distinct coherent states,
often called coherent-state superposition (CSS), 
is regarded as a realization of 
Schr\"{o}dinger's famous cat paradox. 
Typical CSS states are defined as 
$\ket{C_\pm}
=\frac{1}{\sqrt{N_{\pm}}}(\ket{\alpha}\pm\ket{-\alpha})$, 
where $|\pm\alpha\rangle$ are coherent states with amplitudes 
$\pm\alpha$. 
The $\ket{C_+}$ ($\ket{C_-}$) is called an even- (odd-) CSS state 
and $|\alpha|^2$ is often regarded as its ``size'' since it reflects 
the distance of two superposed coherent states. 
Such states have been realized in a few physical systems \cite{Monroe, Brune}. 
Among them, large size CSS states of travelling light are important 
for many quantum information tasks such as 
linear-optics quantum computation \cite{Ralph, Lund}, 
quantum teleportation \cite{vanEnk}, 
and quantum metrology \cite{Gilchrist}. 
In practice, one can generate approximate $\ket{C_\pm}$ 
by subtracting photons from a squeezed vacuum 
\cite{Dakna}. 
Along this line, 
single-photon subtraction has been demonstrated, 
generating odd cat-like states 
\cite{Alex1, Jonas, Wakui}.  
But their sizes ware still small, 
typically $\abs{\alpha}^2\simeq1.0$.


Although subtracting more than two photons leads to 
larger states, 
experiments get more challenging 
due to rapid decrease of the success probability. 
Recently an alternative way utilizing a photon-number state and homodyne detection 
was proposed and a generation of the state $\sqrt{2/3}\ket{2}-\sqrt{1/3}\ket{0}$ 
from a pulsed two-photon state was demonstrated
\cite{Alex2}. 
It is close to a superposition of two squeezed states 
where they are displaced from the origin in opposite directions by the amount $\simeq1.2$ in the phase space. 
If 3.5\,dB squeezing is further applied, 
the state would be $\ket{C_{+}}$ with
$\abs{\alpha}^2\simeq2.6$. 
However, it remains a challenge
to apply squeezing onto such a squeezed CSS state.

In this letter, we propose and demonstrate a novel way 
to enhance the size of CSS states 
without resorting to further subtraction of more than two photons 
or squeezing operations. 
Instead we introduce an ancillary mode 
to assist in suppressing the weights of 
smaller number photons in a CSS state, 
and hence to enhance its size. 
In the experiment, we implement it in the time domain, 
by time-separated two-photon subtraction from continuous wave (cw) squeezed light.

Photon subtraction is done 
by tapping a small fraction of the squeezed vacuum 
for photon counting, 
and by selecting the transmitted state 
conditioned on the detection of photons. 
A two-photon subtraction from a single-mode 
squeezed vacuum (without ancillae) is described as 
\begin{equation}
\label{single_mode_TPS}
\hat{a}^2 \hat{S}(\eps_0)\ket{0} 
=\beta_0\, \hat{S}(\eps_0) \left( \eps_0 
\hat{a}^{\dagger \, 2} + 1 \right) \ket{0} , 
\end{equation}
where $\hat{a}$ is an annihilation operator 
describing one-photon subtraction, $\hat{S}$ is a squeezing operator,   
$\eps_0$ represents the degree of squeezing \cite{comment1} and $\beta_0=\eps_0/(1-\eps_0^2)$. 
This is a squeezed state of the superposition 
$\sqrt{2}\,\eps_0\ket{2}+\ket{0}$, 
which well approximates an even CSS state with the size 
up to $\abs{\alpha}^2\simeq 1.0$ \cite{Dakna}. 
However, it is still possible to further increase the size if 
one could optimize the ratio between $\ket{0}$ and $\ket{2}$ in the
superposition, independently from $\eps_0$\cite{Takeoka}.

\begin{figure}
[thbp]
\begin{center}
\includegraphics[width=1.0\linewidth]{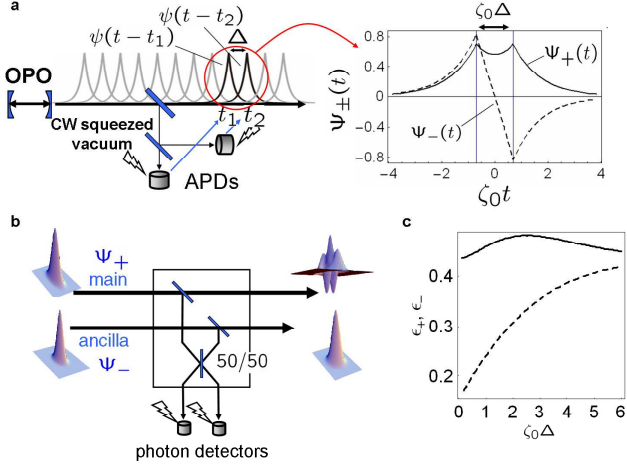}
\caption{(color online) 
(a) The time-separated two-photon subtraction 
and the temporal mode functions $\Psi_\pm$(t). 
OPO: optical parametric oscillator, APD: avalanche photo diode.
(b) Schematic of the ancilla-assisted photon subtraction.
(c) $\Delta$ dependence of the squeezing $\eps_+$ (solid line) and $\eps_-$ (dashed line). $\epsilon = 0.3$. 
}
\label{fig:modeFunc}
\end{center}
\end{figure}

\begin{figure}
\begin{center}
\includegraphics[width=1\linewidth]{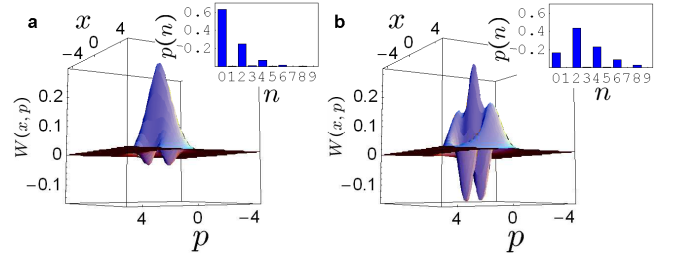}
\caption{(color online) 
Theoretical Wigner functions of the CSS state generated by 
the time-separated two-photon subtraction and their photon number distributions. $\eps=0.3$. 
(a) $\zeta_0 \Delta=0$ and (b) $\zeta_0 \Delta=1.4$ 
correspond to approximate $\ket{C_{+}}$ of 
$\abs{\alpha}^2=1.2$ and $2.6$, respectively. 
}
\label{fig:idealCat} 
\end{center}
\end{figure}

Suppose two independent photon-subtraction events instantaneously occur 
at time $t_1$ and $t_2$, respectively, on a cw squeezed vacuum 
generated from an optical parametric oscillator (OPO) 
as depicted in \reffig{fig:modeFunc}(a). 
Since the cw squeezed vacuum has a finite bandwidth $\zeta_0$, 
the effect of each photon subtraction  
spreads over the temporal wavepacket $\psi(t-t_{1,2}) = \sqrt{\zeta_0} 
e^{-\zeta_0|t-t_{1,2}|}$ in the transmitted beam 
\cite{Molmer,Nielsen1,Nielsen2}. 
When the time separation $\Delta \equiv |t_2-t_1|$ satisfies $\zeta_0 \Delta \lesssim 1$, these two packets are highly overlapped and two temporal waveforms defined as $\Psi_\pm (t) \equiv [\psi(t-t_1)\pm\psi(t-t_2)]/\sqrt{2(1 \pm I_\Delta)}$ with $I_\Delta \equiv (1+\zeta_0\Delta) e^{-\zeta_0\Delta}$  
appear as orthonormal mode functions to describe the process \cite{Nielsen2, Takeoka}. 
The symmetric mode $\Psi_+(t)$ acts as the main mode to be measured while the asymmetric mode $\Psi_-(t)$ serves as an ancilla and
the essential mechanism can be well described by these two modes.
Precisely speaking, subtracted photons have finite correlations with the other temporal modes in the transmitted beam, but the effects of those modes can simply be taken as a small optical loss \cite{Takeoka}.
Then the temporal two-mode model can be translated into a simplified spatial two-mode model with modes $\Psi_\pm$ and a 50/50 beam splitter (BS) as illustrated in \reffig{fig:modeFunc}(b).
The coincidence of clicks at each photon detector always means that two photons must come from either $\Psi_+$ or $\Psi_-$, and not that single photon from each mode.
This is due to the bunching nature of photons.
These two subtraction processes are indistinguishable, producing a superposition, 
\begin{equation}
\label{two_mode_TPS}
 \left( \hat{a}_+^2 - \hat{a}_-^2 \right) 
 \hat{S}_+(\eps_+) \hat{S}_-(\eps_-) \ket{0}_+ \ket{0}_- ,
\end{equation}
where the subscripts $\pm$ indicate modes $\Psi_\pm$. 
The $\eps_\pm$ represent effective degrees of squeezing in each mode 
which are determined by a fully multimode theory and depend on $\Delta$, $\zeta_0$, and 
the OPO pumping parameter $\epsilon$ \cite{Takeoka}.
Figure \ref{fig:modeFunc}(c) shows typical behaviours of $\epsilon_\pm$ 
as functions of $\Delta$.
Since $\epsilon_-$ is relatively small at $\zeta_0 \Delta \lesssim 1$, using \refeq{single_mode_TPS}, 
\refeq{two_mode_TPS} can be rewritten as follows by neglecting terms proportional to $\eps_-^2$ or higher orders,
\begin{align}
\label{two_mode_TPS_re}
&\mathrm{\refeq{two_mode_TPS}} \approx\nonumber\\
&\beta_+\hat{S}_+(\eps_+) \left(\eps_+ \hat{a}_+^{\dagger \, 2} + \left(1 - \frac{\eps_-}{\beta_+}\right) \right)\ket{0}_+ \hat{S}_-(\eps_-) \ket{0}_- 
\nonumber \\
&\equiv\mathcal{N}\ket{\Phi}_+\hat{S}_-(\eps_-) \ket{0}_-,
\end{align}
where $\beta_+=\eps_+/(1-\eps_+^2)$, a normalized state $\ket{\Phi}$ and a normalization factor $\mathcal{N}$  are introduced.
Compared with \refeq{single_mode_TPS}, it can be seen that the ancilla mode provides additional controllability
for tuning  the superposition in the main mode through $\eps_-$, which is simply controlled by changing 
the time separation $\Delta$ as shown in \reffig{fig:modeFunc}(c). This allows us to coherently modify the photon number distribution of the state in the main mode and in particular suppress the weight of small photon number components to increase the size of the CSS state.
Note that, in a rigorous cw model, 
the state in the main mode is slightly degraded to a statistical mixture 
$\hat{\rho}_+ = (1-C_0) |\Phi\rangle\langle\Phi| 
+ C_0 \hat{S}_+ |0\rangle\langle0| \hat{S}_+^\dagger$ 
even without practical imperfections 
due to a weak entanglement between the main and ancilla mode. 
Figure \ref{fig:idealCat} compares the theoretical Wigner functions 
of $\hat{\rho}_+$ based on the full cw model \cite{Takeoka} for $\zeta_0 
\Delta=0$ and $1.4$, i.e. without and with the ancilla assistance. 
These correspond to the CSS states of $|\alpha|^2 = 1.2$ ($\zeta_0 \Delta=0$) and 2.6 ($\zeta_0 \Delta=1.4$) 
with the fidelities of 0.928 and 0.932, respectively,
where the latter $\Delta$ is chosen to maximize the size of CSS while preserving a high fidelity ($>0.9$). 
For larger $\Delta$, $C_0$ becomes non-negligible as discussed later.

\begin{figure}
\begin{center}
\includegraphics[width=0.9\linewidth]{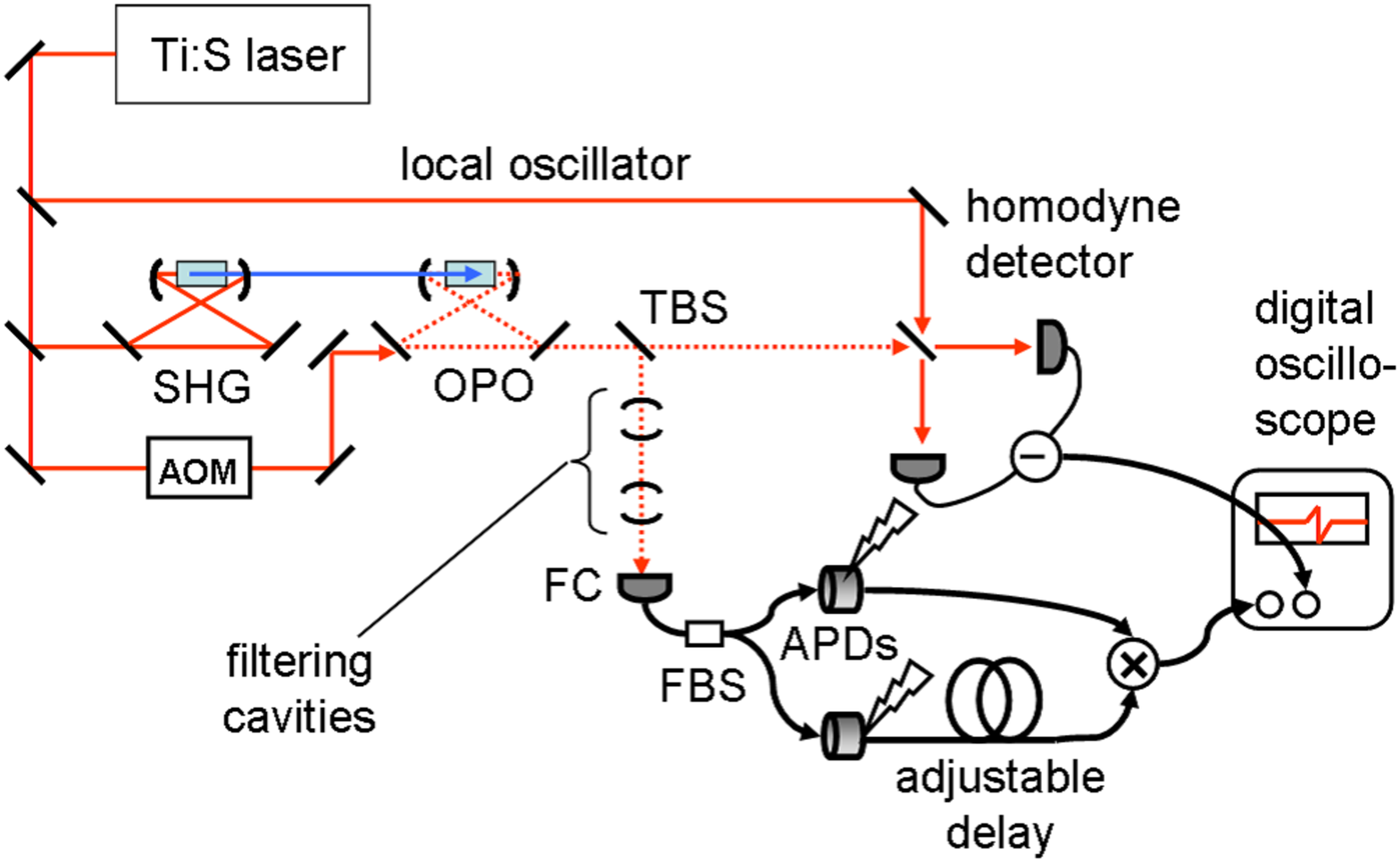}
\caption{(color online) 
Experimental setup of the time-separated two-photon 
subtraction. 
SHG : second harmonic generation cavity, 
OPO : optical parametric oscillator, 
AOM : acousto-optic modulator, 
FC : fiber coupler, 
TBS : tapping beam splitter ($R=$10\%)
FBS : 50/50 fiber beam splitter, and 
APDs : Si avalanche photo diodes (Perkin-Elmer SPCM-AQR-14). 
}
\label{fig:setup}
\end{center}
\end{figure}

A Schematic of our experimental setup is shown in 
\reffig{fig:setup}. 
It is built on our previous single-photon subtraction setup \cite{Wakui} 
with qualitative updates to cope with long data acquisition for hours. 
A cw Ti:Sapphire laser of 860nm is used as a light source. 
A squeezed beam is generated from an OPO which employs 
a periodically-poled KTiOPO$_4$ (PPKTP) as nonlinear crystal. 
The OPO is pumped by a frequency doubled beam of 20mW 
from a second harmonic generation (SHG) cavity with a KNbO$_3$ crystal. 
This corresponds to $\epsilon = 0.3$. 
The OPO bandwidth is $\zeta_0/2\pi\sim$4.5\,MHz (HWHM). 
A small fraction (10\%) of the beam 
is tapped by a tapping BS (TBS) and going through two filtering 
Fabry-Perot cavities of 2mm-long and of 0.9mm-long, respectively. 
The unwanted photons in non-degenerate modes extending over a wide range of 
frequencies are well filtered out by these cavities and 
only the component in the degenerate mode around 860nm is 
guided into two avalanche photo diodes (APDs) \cite{Jonas, Wakui}. 

The beam transmitted through the TBS 
is continuously measured by a homodyne detector. 
All cavities are actively locked on the resonance for 860nm 
by a weak coherent beam and electronic feedback. 
The coherent beam 
is periodically chopped by an acousto-optic modulator (AOM), 
which defines distinct time bins by presence 
and absence of the coherent beam, 
and thus enables the alternate sequence of locking and 
photon counting without stray light. 
Also via phase locking the coherent beam 
to parametric gain of the OPO, 
we obtain phase information of the measured state 
from interference between the coherent beam and the local oscillator (LO) 
beam at the homodyne detector. 

Clicks of APDs announce that a conditional state appears in the transmitted beam. 
The time separation $\Delta$ is adjusted by an electronic delay line. 
The pumping of $\eps=0.3$ at the OPO outputs a 3.0dB squeezed 
vacuum in temporal mode $\Psi_+(t)$ and 
the simultaneous-click rate of the two APDs is about 6 counts per second for $\Delta=0$ns and less than 1 count per second
for $\Delta \sim 100$ns. 
The conditional state in the main mode is measured by a homodyne detector to reconstruct its Wigner function. 
For every pair of clicks of the APDs with desired $\Delta$, a continuous photo-current of the homodyne detector 
is stored. 
A quadrature amplitude in 
the main mode is extracted from the continuous data by 
applying the numerical mode filter with the shape of $\Psi_+(t)$ 
which is equivalent to an optical LO with the waveform of $\Psi_+(t)$ 
\cite{Jonas,Wakui}.

\begin{figure*}
\begin{center}
\includegraphics[width=0.8\linewidth]{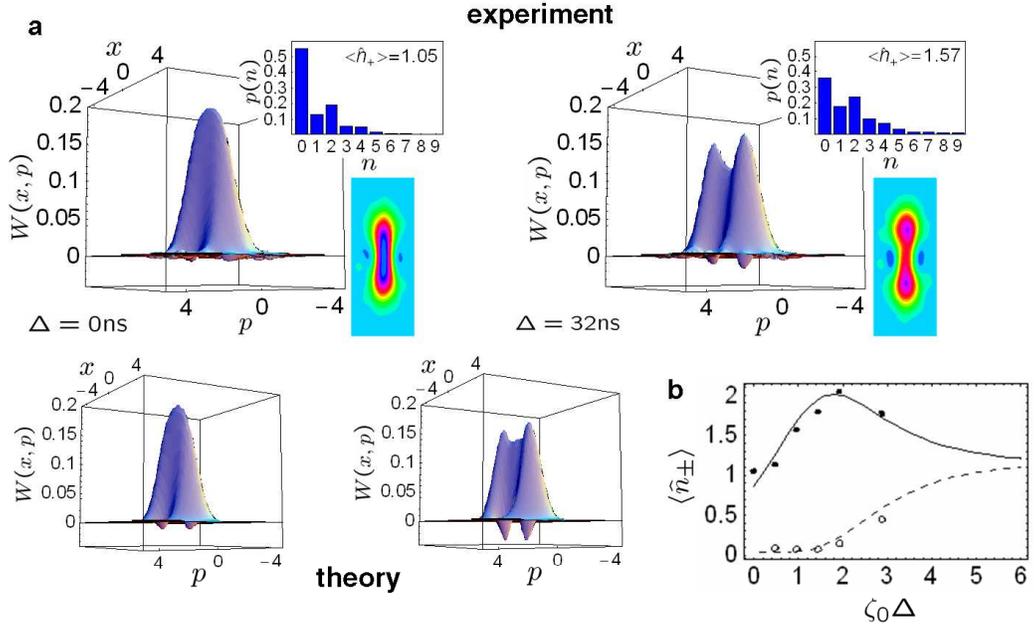}
\caption{(color online) 
(a) Wigner functions of the state 
$\hat \rho_+$ in mode $\Psi_{+}$ for $\Delta=0$ns ($\zeta_0 \Delta=0$, left) 
and $\Delta=32$ns ($\zeta_0 \Delta=0.90$, right). 
Upper row: experimental Wigner functions ($W(x,p)$), 
their contours, and photon number distributions ($p(n)$). 
Lower row: model calculation of the Wigner functions.
(b) Average photon numbers in modes $\Psi_\pm$. 
Upper (filled) plots: experimental $\langle \hat{n}_+ \rangle$ 
with $\Delta =0$, 32, 48, 64, 96 ns, from left to right. 
Lower (unfilled) plots: experimental $\langle \hat{n}_- \rangle$ 
with $\Delta =32$, 48, 64, 96 ns, from left to right. 
Solid ($\langle \hat{n}_+ \rangle$) 
and dashed ($\langle \hat{n}_- \rangle$) 
lines are the theoretical predictions 
including 15\% optical loss. 
}
\label{fig:catDelta}
\end{center}
\end{figure*}

Experimentally reconstructed Wigner functions and photon number 
distributions of $\hat{\rho}_+$ with $\Delta=0$ and 32ns 
are given in the upper row of \reffig{fig:catDelta}(a). 
The reconstruction was performed by the conventional maximum likelihood
estimation method \cite{Lvovsky} from 22,000 points 
of quadrature amplitudes in the raw experimental data. 
No experimental imperfections are corrected in the reconstruction 
procedure, 
whereas most of the previous works corrected 
the losses at the state evaluation step 
\cite{Alex1,Jonas,Alex2}. 
Figure \ref{fig:catDelta}(a) clearly shows that 
with finite $\Delta$, 
the two positive peaks corresponding to the components 
$\ket{\pm\alpha}$ become apart, 
i.e. the size-increase of $\alpha$.

Ideally, 
the two-photon subtracted states of $\Delta=$0 and 32ns 
($\zeta_0 \Delta=0$ and 0.9, respectively) 
would approximate $\ket{C_+}$ of 
$\abs{\alpha}^2=$1.2 and 2.0, respectively, 
with high fidelities ($>90\%$). 
In the experiment, the fidelities of 
76\% and 60\%, for $\abs{\alpha}^2=$1.2 and 2.0, respectively,  
are achieved. 
A particular feature of the macroscopic quantum superposition is 
the presence of negative dips in the Wigner function 
due to the interference between two distinct wavepackets. 
For $\Delta=$32\,ns, 
the negative dips are clearly observed at the sides of the origin. 
This is, to our knowledge, the largest CSS state of the travelling light ever observed, 
showing the negative Wigner function without any compensation of 
experimental imperfections. 
Such quality is achieved by use of 
a highly pure squeezed vacuum 
generated from our low loss OPO\,\cite{Wakui}. 
The sources of imperfection are separately investigated. 
The OPO has 0.4\% intra-cavity loss. 
Associated with 9.5\% transmittance of the coupling mirror, 
it gives 0.95 of the escape efficiency. 
The propagation efficiency from the OPO to the homodyne detector 
is 0.95. 
The quantum efficiency of the homodyne detector and its electronic noise give effective detection efficiency 0.98. 
The spatial overlap between the squeezed vacuum and the LO gives 
another efficiency of 0.96. 
In total, 
the overall efficiency 
is estimated around 0.85.
Theoretical Wigner functions 
including these imperfections 
are shown in the lower row of \reffig{fig:catDelta}(a). 
The experimental results are almost in agreement with 
the theoretical models \cite{Sasaki2}
while the negativity is slightly degraded by some uncleared imperfections. 
We also note that the maximum CSS state is obtained around $\Delta=48$ns 
(see \reffig{fig:idealCat}(b)) 
while the experimentally observed negative dips are limited up to 32ns 
since the low count rates at large $\Delta$ also degrades the negativity.

The photon number distribution in \reffig{fig:catDelta}(a) 
also reveals a mechanism of the ancilla-assisted photon-subtraction.
Suppression of the small photon-number components by increasing $\Delta$ 
is clearly observed. 
Our result should also be contrasted to the squeezed state superposition 
$\sqrt{2/3}\ket{2}-\sqrt{1/3}\ket{0}$ 
generated in \cite{Alex2}. 
In our case, 
the even-number distribution extends at least up to $n=4$, 
and is clearly deviated from that of the squeezed vacuum, 
which should be another signature of 
the macroscopic nature of the superposition.

Figure\,\ref{fig:catDelta}(b) plots the average photon numbers 
$\langle\hat n_\pm\rangle$ of modes 
$\Psi_{\pm}$ as a function of $\Delta$, 
that agree with the theory well indicated by lines \cite{Sasaki2}. 
The increase of $\langle \hat{n}_+ \rangle$ 
in the range $\zeta_0\Delta\alt2$ supports 
an experimental evidence of the size-increase. 
In this range most of the photons accumulate in the main mode with 
fewer photons in the ancillary mode. 
As $\Delta$ increases beyond $\zeta_0\Delta\agt2$, 
however, 
$\langle\hat n_-\rangle$ starts to increase, 
and photons in modes $\Psi_+$ and $\Psi_-$ are getting entangled. 
It induces degradation of the purity of the reduced state in the main mode. 
For $\zeta_0\Delta\gg1$, 
the state in the main mode is degraded to a completely mixed 
state $\hat{\rho}_+ \to \frac{1}{2} \hat{S}_+ 
( |2\rangle\langle2| + |0\rangle\langle0| ) \hat{S}_+^\dagger$.

Finally, 
we point out that the above drawback of the purity degradation  
could be overcome 
if one would take advantage of an ancillary mode 
which is not  entangled to the main mode. 
This is done by preparing a coherent state 
$\ket{\alpha}$ in the ancillary mode. 
The improved scheme is represented 
in terms of generic two modes A and B as 
\begin{align}
(\ani_A^2-\ani_B^2)\hat{S}_A \ket{0}_A\ket{\alpha}_B
=(\ani_A^2-\alpha^2)\hat{S}_A \ket{0}_A\ket{\alpha}_B. 
\end{align}
Since the coherent state is an eigenstate of $\hat{a}$, 
one can make a superposition of two- and zero-photon subtractions 
in mode $A$ without entangling two modes, i.e. without degradation. 
Here the coherent amplitude simply plays the role of $\Delta$ 
in the time-separated two-photon subtraction.
In practice, the experiment will get more involved since one has to lock the relative phase of
 these two modes. 
Also a related scheme 
was proposed recently \cite{Nielsen3}.

In conclusion, we have proposed 
a concept of the ancilla-assisted photon-subtraction
and experimentally demonstrated it by 
the time-separated two-photon subtraction from cw squeezed light. 
Due to the quantum interference assisted by the ancillary mode, 
we have successfully observed and characterized the largest CSS state of travelling light 
so far showing 
the negative dips of the Wigner function without 
any corrections of measurement imperfections. 
Our experimental scheme involves the trade-off 
between the size-enlargement and the degradation of the state purity. 
We pointed out that the trade-off can be circumvented
by extending it to the one with a coherent-state ancilla. 


This work was supported by 
a MEXT Grant-in-Aid for Scientific Research (B) No.~19340115, 
and for Young Scientists (B) No.~19740253.



\begin{references}
\bibitem{comment1}
With the conventional squeezing parameter $r$, the squeezing operator is defined as $\hat{S}(r)=\exp[\frac{r}{2}\left(\hat{a}^2-\hat{a}^{\dagger \,2}\right)]$.
$\eps_0$ and $r$ are related as $\eps_0 = \tanh r$. 
See Refs~\cite{Takeoka,Sasaki2} and references therein.
\end{references}
\end{document}